\documentclass[journal]{IEEEtran}
\usepackage{graphicx}
\ifCLASSINFOpdf
\else
\fi
\usepackage[cmex10]{amsmath}
\usepackage{amsmath,amssymb,amsthm}
\usepackage{amsfonts}

\begin{document}
%
\title{How Much Power Must We Extract From a Receiver Antenna to Effect Communications?}
%
%
%

\author{Thomas~L.~Marzetta,~\IEEEmembership{Life~Fellow,~IEEE,}
        Brian~McMinn,~\IEEEmembership{Student Member,~IEEE,} Amritpal~Singh,~\IEEEmembership{Student Member,~IEEE,} and~Thorkild~B.~Hansen,~\IEEEmembership{Senior~Member,~IEEE}
\thanks{T. L. Marzetta, B. McMinn, and A. Singh are with the Electrical and Computer Engineering Department, New York University.}
\thanks{T. B. Hansen is with Seknion, Inc.}
\thanks{Research of T. L. Marzetta, B. McMinn, and A. Singh was supported by NYU WIRELESS.}}

%
%

\markboth{IEEE Journal on Selected Areas in Information Theory: Special Issue on Electromagnetic Information Theory}%
{Shell \MakeLowercase{\textit{et al.}}: Bare Demo of IEEEtran.cls for Journals}
%



\maketitle

\begin{abstract}
Subject to the laws of classical physics - the science that governs the design of today's wireless communication systems - there is no need to extract power from a receiver antenna in order to effect communications. If we dispense with a transmission line and, instead, make the front-end electronics colocated with the antenna, then a high input-impedance preamplifier can measure the open-circuit voltage directly on the antenna port without drawing either current or power. Neither Friis' concept of noise figure, nor Shannon information theory, nor electronics technology dictates that we must extract power from an antenna. 
\end{abstract}


%
\IEEEpeerreviewmaketitle

\section{Introduction}
%
%
%
%
\IEEEPARstart{T}{his} paper addresses the question whether or not it is necessary, from the perspective of classical physics, to extract power from a receiver antenna for the purpose of communication. Remarkably, this fundamental question has not been posed in the literature: apparently nobody says that power extraction is essential; conversely nobody says that power extraction is unnecessary.

\subsection{Classical Statistical Mechanics}
We address this question entirely in the context of classical physics. While the solid-state physics used to design semiconductor devices is grounded in quantum mechanics, all of today's cellular wireless systems are designed according to classical physics. 
Bra-ket notation has no role in present-day wireless engineering. No wireless engineer uses Schroedinger's equation as a design tool.

\subsection{Why Would Anyone Think We Have to Extract Power From a Receiver Antenna?}
Several widely known facts seem to suggest the necessity of power extraction:

\begin{itemize}

\item Most wireless receivers utilize a transmission line to connect the antenna to the active electronics. The transmission line is impedance-matched to the antenna, and terminated in a resistance. Hence power is extracted from the antenna.

\item The Friis definition of amplifier gain is the available power at the output of the amplifier divided by the available power at the input. This suggests that the amplifier is driven by power at its input which can only come from the antenna.

\item Shannon information theory produced a famous formula, $E_{\mathrm{b}}/N_0 > \ln 2$. The quantity, $E_{\mathrm{b}}$, is commonly referred to as ``energy per bit", which, if taken literally, implies that the energy must come from the antenna.

\end{itemize}

\subsection{Why We Do Not Have to Extract Power From a Receiver Antenna}
This paper systematically refutes the notion that we have to extract power from a receiver antenna:
\begin{itemize}

\item Section II argues that Shannon information theory is a purely mathematical theory, which \emph{per se} tells us nothing about the physical world.

\item Section III presents amplifier circuits capable of measuring the open-circuit voltage on an antenna port.

\item Section IV contrasts the implications of connecting the antenna to the amplifier with a transmission line versus making the electronics colocated with the antenna. Impedance-matching is not always optimal.

\item Section V establishes that the concept of noise figure is solely a standards definition, useful because it ensures that 50-Ohm microwave devices can be connected together with 50-Ohm coaxial cable with predictable results.  Noise figure, however, embodies no fundamental physical principle.

\item Section VI shows that impedance matching, and power extraction, can sometimes be used advantageously to obtain noise-free voltage gain prior to the preamplifier.

\item Section VII applies lessons learned to broad-band receiver antenna arrays, and suggests possible advantages of simply measuring open-circuit voltages on the antenna ports as an alternative to drawing current from the antennas.
    
\end{itemize}

\section{Does Shannon Information Theory Provide Any Guidance?}
Shannon theory \cite{Shannon1} is a mathematical theory involving non-dimensional statistical quantities which can be thought of as pure numbers residing in a digital computer. The square of any quantity is handily called ``power", which having no dimensions has no physical significance. Likewise ``signal-to-noise" ratio is a ratio of the variances of two non-dimensional independent random variables, with the numerator associated with information and the denominator associated with degradation of communications.

In order to use Shannon theory to analyze a wireless communication system, we need a model of how an antenna interacts with the incoming electromagnetic field, and in turn, how the voltage generated at the antenna port, along with noise, finds its way via passive circuits, amplifiers, mixers, and A/D converters, into a digital computer. For this model we depend entirely on the physics of antennas and the theory of electrical circuits. Shannon theory, itself, provides no guidance.

\subsection{Band-Limited Additive Complex Gaussian Noise Channel}
Under the complex base-band representation, $x(t)$ is the continuous-valued transmit signal, $y(t)$ is the continuous-valued receive signal, and $t$ is a continuous-valued independent variable,
\begin{equation} \label{add_Gauss_channel}
    y(t) = x(t) + w(t) ,
\end{equation}
where $w(t)$ is statistically independent white Gaussian noise.
All quantities are dimensionless, but $t$ can be conveniently regarded as "time".\footnote{We could equally well regard $t$ as spatial position.} The transmitted signal is constrained in both bandwidth and expected ``power": $\omega \in [- \pi B, \pi B]$, $\mathbb{E} |x(t)|^2 \leq P$, where $P$ and $B$ are ``power" and ``bandwidth" respectively. The noise, $w(t)$, is zero-mean circularly-symmetric complex Gaussian with spectral density $N_0$. The variance of the band-limited noise is $\mathbb{E} |w(t)|^2 = N_0 B$. Therefore the SNR of the received signal is $\mathrm{SNR} = P/N_0 B$.

\subsection{Channel Capacity and ``Energy" per Bit}
The capacity of the channel, in units of bits per ``second", is \cite{Shannon2}
\begin{equation} \label{cap_gauss}
    C = B \log_2 \left( 1 + \frac{P}{B N_0} \right) .
\end{equation}
For fixed transmit ``power", the capacity increases monotonically with increasing ``bandwidth". The infinite-bandwidth capacity constitutes an upper bound,
\begin{equation} \label{cap_gauss_lim}
C < \frac{P}{N_0 \ln 2} .
\end{equation}
The ``time" required to transmit a single bit is equal to $1/C$, entailing a transmit ``energy"-per-bit of $E_{\mathrm{b}} = P/C$. With this notation we can rewrite (\ref{cap_gauss_lim}) as the celebrated formula,
\begin{equation} \label{EbN0}
    \frac{E_{\mathrm{b}}}{N_0} > \ln 2 .
\end{equation}

\subsection{Comments}
The inequality (\ref{EbN0}) is absolutely correct, mathematically. But a superficial reading of Shannon's paper might lead to a gratuitous physical interpretation for the quantity $E_{\mathrm{b}}$: that is represents the physical energy per bit that has to be extracted from the antenna. Such an interpretation is completely unwarranted, nor is it implied anywhere in Shannon's paper.

\subsection{Take-Away Points}
\begin{itemize}
    \item Shannon information theory is the standard tool for the theoretical analysis of wireless communication system.
    \item Shannon theory is purely mathematical rather than physical. In particular it describes statistically-based activities involving non-dimensional quantities. Effectively, everything takes place inside a digital computer.
    \item Shannon theory does not tell us how the electromagnetic field that is sensed by the antenna is related to the non-dimensional quantities in the computer. This can only be modelled by an application of electromagnetic theory and circuit theory.
    \item The formula $E_{\mathrm{b}}/N_0 > \ln 2$ has no automatic physical interpretation and Shannon theory alone does not imply that $E_{\mathrm{b}}$ represents physical energy extracted from an antenna.
\end{itemize}

\section{Can We Measure the Open-Circuit Voltage on an Antenna?}
\begin{figure}
\centering
\includegraphics [scale=0.7] {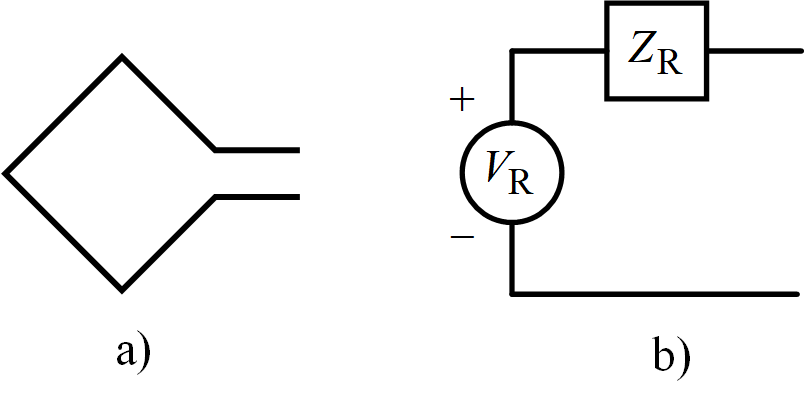}
\caption{Receiver loop antenna a) and Thevenin-equivalent circuit b): the open-circuit voltage is proportional to the time-derivative of the magnetic flux passing through the loop, and the series impedance is equal to the self-impedance of the antenna.}
\label{antenna}
\end{figure}

A small loop antenna is physically modelled as an ideal voltage source in series with a complex-valued impedance, Fig. \ref{antenna}.\footnote{We assume a loop antenna which provides a DC path for bias currents associated with preamplifiers. In contrast a dipole would require a transformer connection to provide a DC path.} The open-circuit voltage is proportional to the time-derivative of the magnetic flux  passing through the loop. The series impedance is equal to the self-impedance of the antenna. The electromagnetic field is the superposition of the desired signal, man-made interference, naturally-occurring interference, including static and background (dark-sky) noise, and electromagnetic scattering with the antenna itself. The antenna is assumed lossless. Having no internal resistance, it is therefore not a source of Johnson noise \cite{Pierce}.

\begin{figure}
\centering
\includegraphics [scale=0.7] {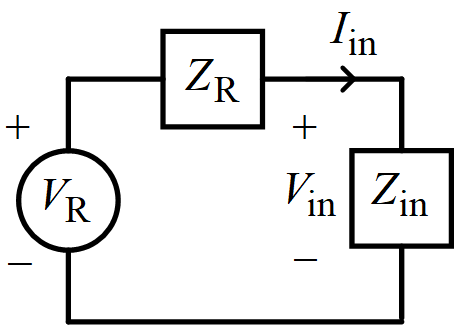}
\caption{Receiver antenna is connected to a preamplifier having input impedance, $Z_{\mathrm{in}}$.}
\label{ant_load}
\end{figure}

Suppose that the antenna is connected directly (with no intervening transmission line) to a preamplifier having complex input impedance, $Z_{\mathrm{in}} = Z_{\mathrm{in}}' + i Z_{\mathrm{in}}''$, Fig. \ref{ant_load}.
The power extracted from the antenna is equal to
\begin{equation} \label{ant_power}
P = \frac{|V_{\mathrm{R}}|^2 \,  Z_{\mathrm{in}}'}{2 \left[  \left( Z_{\mathrm{R}}' + Z_{\mathrm{in}}' \right)^2   + \left( Z_{\mathrm{R}}'' + Z_{\mathrm{in}}'' \right)^2\right]} .
\end{equation}
Conjugate-matching, $Z_{\mathrm{in}} = Z_{\mathrm{R}}^{*}$, extracts maximum power, $P_{\mathrm{max}} = |V_{\mathrm{R}}|^2 /  8 Z_{\mathrm{R}}'$. 
For a very high preamplifier real input impedance, $Z_{\mathrm{in}}' \gg Z_{\mathrm{R}}'$, 
the ratio of power extracted to maximum available power is bounded as $P / P_{\mathrm{max}} < Z_{\mathrm{R}}' /Z_{\mathrm{in}}'$.
Thus the real part of the amplifier input impedance should be as large as possible to minimize power extraction from the antenna.

Another undesirable effect of finite input impedance, $Z_{\mathrm{in}}$, is to reduce the input voltage to the amplifier through the voltage divider effect,
\begin{equation} \label{v_in_load}
V_{\mathrm{in}} = \frac{V_{\mathrm{R}} Z_{\mathrm{in}}}{Z_{\mathrm{R}} + Z_{\mathrm{in}}} .
\end{equation}
For example, impedance-matching can reduce the magnitude of the measured voltage by as much as a factor of two. When the input impedance comprises a resistor in parallel with a capacitor, then the capacitor can present a low reactive impedance at sufficiently high frequencies thereby degrading the signal; in theory, however, this low impedance can be cancelled at a single frequency by a shunt inductance, and over a wide band by a more elaborate passive network.

In the following we discuss two basic methods for measuring the open-circuit voltage across the antenna port. The first employs a preamplifier having inherently high input impedance with respect to the antenna self-impedance. The second is a null-balance measurement whereby the voltage appearing on the antenna port is balanced against a variable reference voltage; at the null point no current is drawn from the antenna.

\subsection{High Input Impedance Pre-Amplifiers}

\begin{figure}
\centering
\includegraphics [scale=0.5] {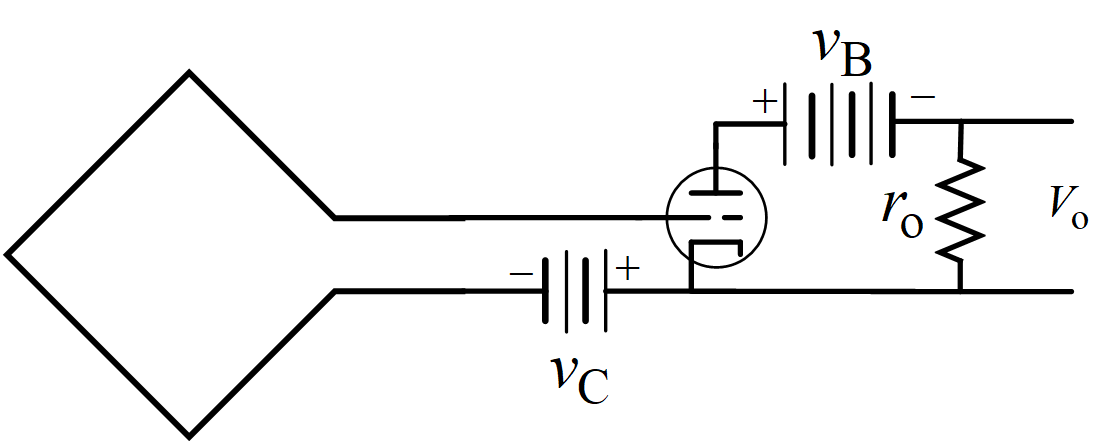}
\caption{Common cathode triode vacuum tube amplifier functions as a voltage-controlled current source: C-battery creates negative grid-to-cathode bias resulting in high input impedance; output power is provided entirely by B-battery which creates a positive plate-to-cathode potential.}
\label{triode}
\end{figure}

\begin{figure}
\centering
\includegraphics [scale=0.5] {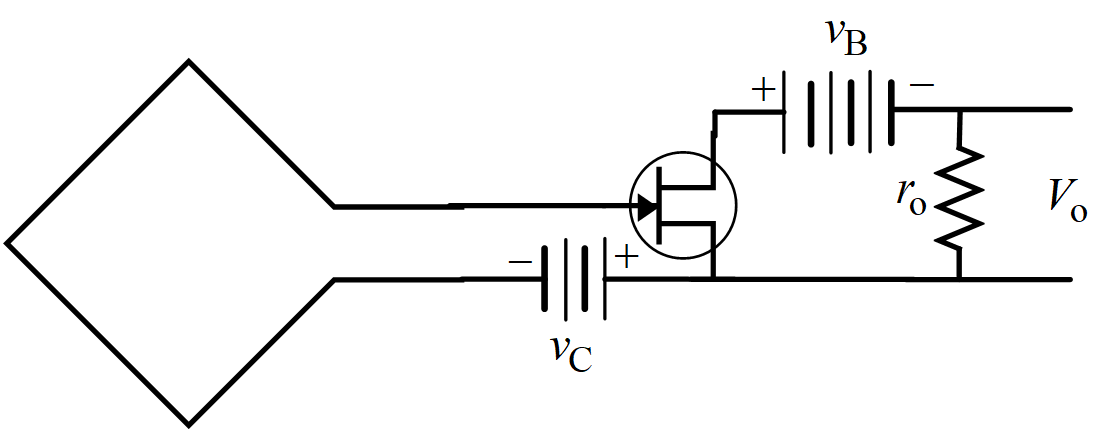}
\caption{Common-source JFET amplifier is the solid-state counterpart of the triode vacuum tube amplifier of Fig. \ref{triode}.}
\label{JFET}
\end{figure}

Figs. \ref{triode} and \ref{JFET} illustrate the classical common-cathode triode vacuum tube amplifier and its modern solid-state counterpart, the common-source N-channel junction field-effect transistor (JFET) amplifier. 
The two amplifying devices are electrically similar.

The gate of the JFET is negatively biased with respect to the source resulting in an extraordinarily high input impedance - equivalent to that of a reverse-biased diode. There is a small d.c. bias current flowing from the source through the loop antenna to the gate. This bias current is incidental to the amplifying action of the JFET, and is not associated with any power extraction from the antenna. It would be a mistake to think that the JFET directly amplifies power; rather it functions as a voltage-controlled current source. The power to drive the load resistor is supplied entirely by the local power supply (the B-battery), not by the antenna.

\begin{figure}
\centering
\includegraphics [scale=0.5] {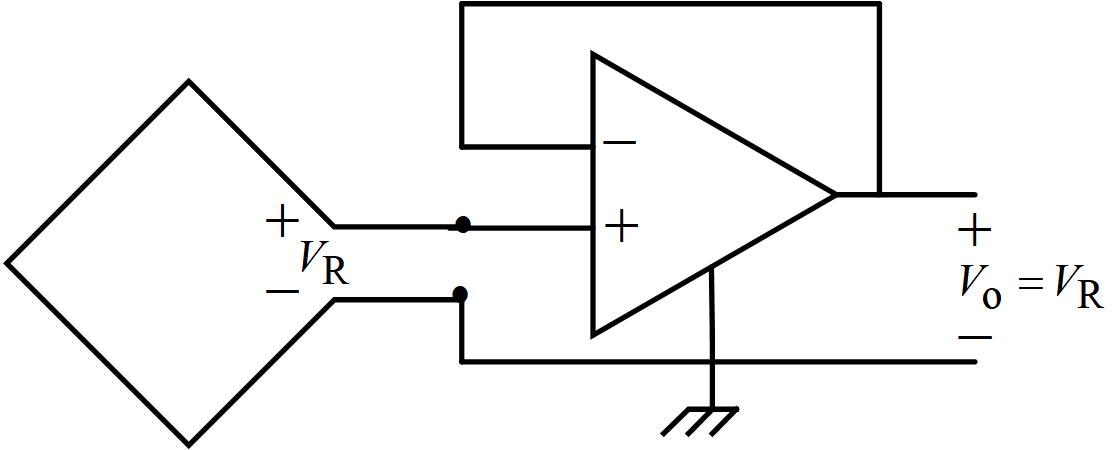}
\caption{Unity-gain buffer amplifier: op amp circuit provides high input impedance and precise voltage-gain of one.}
\label{follower}
\end{figure}

Fig. \ref{follower} illustrates the unity-gain buffer amplifier utilizing an operational amplifier (\emph{op amp}). An op amp presents extremely high impedances at the inverting and non-inverting inputs, and a large open-loop voltage gain. Almost no current is drawn from the antenna. The loop antenna provides a d.c. path for the minute input bias current. The high open-loop gain of the op amp and negative feedback ensure that the inverting input is virtually at the same potential as the non-inverting input, in turn, equal to the open-circuit voltage of the antenna. The direct feedback connection between the output and the inverting input implies that the output voltage is equal to the antenna open-circuit voltage, $V_{\mathrm{R}}$. 

While op amps are not usually considered r.f. devices, they are capable of functioning in the VHF band \cite{EDN_RF_opamp}, \cite{TI_rf_opamp}. The Analog Devices AD8033, for example, has a JFET-input and when used in the unity-gain buffer configuration, has a $- 3$ dB bandwidth of 80 MHz, 92 dB open-loop voltage gain, with an input impedance of $10^{12}$ ohms in parallel with a 2 pF capacitance, and an input bias current of 1 pA \cite{AD8033}. A unity-gain buffer using this op amp would function perfectly well as a preamplifier within an AM radio receiver, e.g., .55 - 1.7 MHz.

The literature concerning r.f. applications of op amps implicitly assumes that the r.f. source is connected to the op amp through a transmission line, which itself must be terminated in a matching resistor. Thus \cite{TI_rf_opamp}, Fig. 4, illustrates the use of a non-inverting op amp, but the input is shunted by a 50-Ohm resistor. The author of this application report evidently overlooked the possibility of dispensing with the transmission line and connecting the op amp directly to the r.f. source without a terminating resistor, thereby presenting a high impedance to the antenna and increasing the voltage gain by a factor of two.

\subsection{Null-Balance Voltage Measurement}

\begin{figure}
\centering
\includegraphics [scale=0.6] {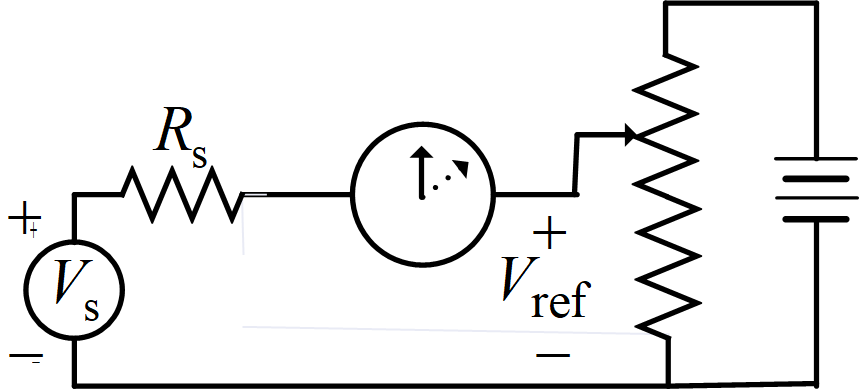}
\caption{Null-balance voltage measurement: the unknown source voltage is balanced against a precision reference voltage; at the null point no current is drawn from the source.}
\label{null_balance}
\end{figure}

Fig. \ref{null_balance} illustrates the classical null-balance d.c. voltage measurement. 
A precision voltage divider of the Kelvin-Varley type generates a variable reference voltage which is balanced against the unknown voltage, as indicated by a sensitive mirror galvanometer. At the null point, the two voltages are equal, $V_{\mathrm{ref}} = V_{\mathrm{s}}$, and no current is drawn from the voltage source.

\begin{figure}
\centering
\includegraphics [scale=0.7] {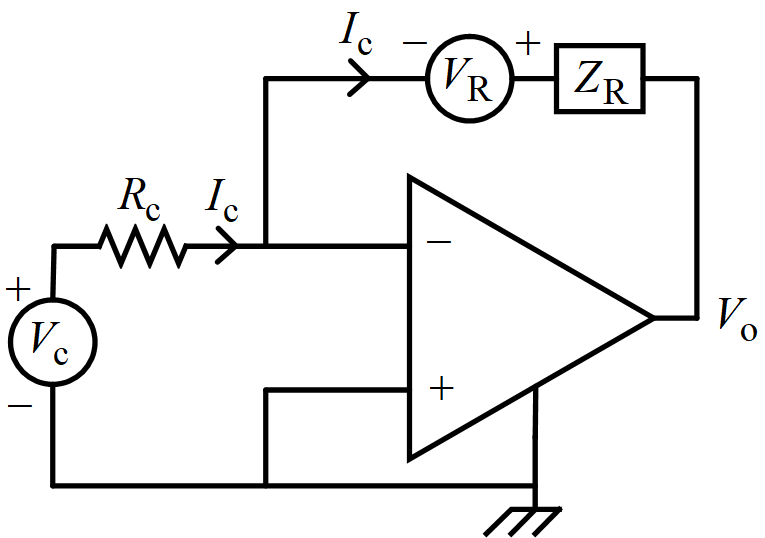}
\caption{Precision constant current source: current, $I_{\mathrm{c}}$, is set by control voltage, $V_{\mathrm{c}}$ and resistor, $R_{\mathrm{c}}$; current $I_{\mathrm{c}} = V_{\mathrm{c}} / R_{\mathrm{c}}$ is independent of the source voltage $V_{\mathrm{R}}$ and impedance $Z_{\mathrm{R}}$.}
\label{constant_current}
\end{figure}

Conceptually, the null-balance technique could measure an r.f. voltage if the voltage-divider were driven by a sufficiently fast servomechanism. The same thing can actually be accomplished with a nearly-forgotten op amp circuit called the \emph{inside-out voltage follower} \cite{Inside_Out}. The circuit is based on the voltage-controlled constant-current source of Fig. \ref{constant_current}.
The inverting input to the op amp draws essentially zero-current because of its high input impedance. The negative feedback and the high open-loop gain ensure that the inverting input is at ground potential. Consequently a current, $I_{\mathrm{c}} = V_{\mathrm{c}}/R_{\mathrm{c}}$, flows through the source which is independent of the source voltage and impedance. The output voltage of the op amp is $V_{\mathrm{o}} = V_{\mathrm{R}} - Z_{\mathrm{R}} I_{\mathrm{c}}$.

\begin{figure}
\centering
\includegraphics [scale=0.6] {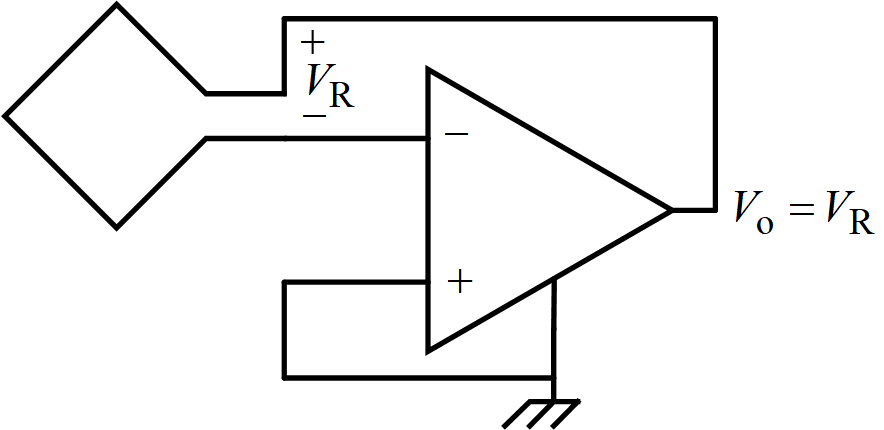}
\caption{Inside-out voltage follower is equivalent to the voltage-controlled current source of Fig. \ref{constant_current} with the current set to zero; negative feedback balances the op amp output voltage against the antenna voltage resulting in zero-current. The input impedance with respect to that of the unity gain buffer of Fig. \ref{follower} is increased by a factor equal to the open-loop gain of the op amp.}
\label{inside_out_ant}
\end{figure}

As shown in Fig. \ref{inside_out_ant}, if we let $V_{\mathrm{c}} = 0$ and $R_\mathrm{c} = \infty$, then the current, $I_{\mathrm{c}}$, is equal to zero and the op amp produces a voltage, which when added to the antenna voltage, produces zero-current: $V_{\mathrm{o}} = V_{\mathrm{R}}$.
The advantage of the inside-out voltage follower over the conventional unity-gain voltage buffer is that the effective input impedance of the former is increased by a factor equal to the open-loop gain of the op amp.

The electronic null-balance measurement of voltage can be extended to the electromagnetic domain \cite{Dolinar}. The Dolinar receiver is a coherent optical homodyne receiver in which the local reference field is driven by the output of the photomultiplier tube detector which itself can detect single photons. Negative feedback drives the reference field so as to cancel the incoming electromagnetic field. Perfect cancellation is not possible because the reference field is driven by the output of the detector, and perfect cancellation would result in a photon arrival rate of zero. Consequently the detector does extract non-zero power, with the energy per photon equal to the product of Planck's constant and the frequency of the photon, $E = h \nu$. However if we neglect quantum physics by pretending that Planck's constant is equal to zero, the Dolinar receiver could deliver a positive bit-rate with an arbitrarily small received power.

\subsection{Take-Away Points}
\begin{itemize}
    \item It is entirely feasible to measure the open-circuit voltage across a receiver antenna port, and in so doing neither current nor power is drawn from the antenna;
    \item The open-circuit voltage can be measured in two ways: 1) a pre-amplifier having an inherently high input impedance, 2) a null-balance measurement;
    \item To measure the open-circuit voltage, implicitly the measuring device can be no more than a fraction of a wavelength from the antenna - otherwise the connecting wires constitute a transmission line with accompanying impedance mismatch effects;
    \item It appears that classical physics provides no upper bound on the effective impedance of the voltage measurement: classically, in order to effect communications, we don't have to extract current - and therefore power - from a receiver antenna.
\end{itemize}

\section{To Match Impedance or Not?}

Typical microwave practice entails connecting antennas and amplifiers together with long (e.g., many-wavelength) transmission lines \cite{Elliott},\cite{Pozar}. To eliminate multiple reflections within a transmission line, it is necessary to terminate the line with an equivalent matching resistance. Much of microwave practice is shaped by the ubiquitous employment of transmission lines. However there is no fundamental requirement to use a transmission line.

\subsection{Simplified Wireless Receiver Front End Employing a Transmission Line}

\begin{figure}
\centering
\includegraphics [scale=0.3] {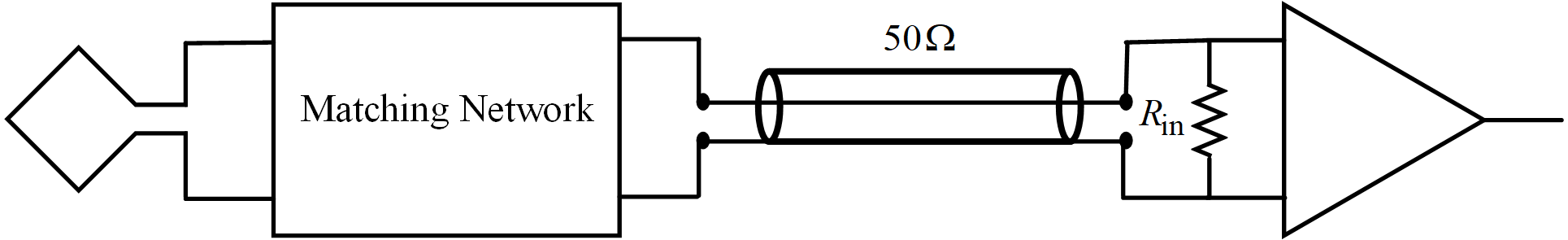}
\caption{Typical receiver front end: Receiver antenna and passive matching network drive a transmission line, terminated in a resistor; the voltage drop on the terminating resistor is measured by a high input impedance preamplifier, but all the power extracted from the antenna is dissipated in the resistor.}
\label{typical_receiver}
\end{figure}

Fig. \ref{typical_receiver} illustrates a simplified diagram of the front end of a representative receiver. 
The antenna and its matching network is connected to the preamplifier by a 50-Ohm transmission line. To avoid reflections the transmission line is terminated in a resistor, $R_{\mathrm{in}}=50 \, \Omega$. \footnote{The 50-Ohm resistor could be replaced by a higher resistance and a passive matching network - sometimes an advantageous thing to do; see Section \ref{Transformer}.} A high input impedance preamplifier measures the voltage on the terminating resistor; the voltage should be as large as possible to defeat internal amplifier noise. In turn this calls for the matching network to extract maximum power from the antenna, all of which is expended in heating the terminating resistor and none of which contributes directly to amplification. In short it is the transmission line - something of no fundamental importance - that drives the design of the front end.

\subsection{Impedance-Matching vs. Open-Circuit Voltage Measurement}

\begin{figure}
\centering
\includegraphics [scale=0.35] {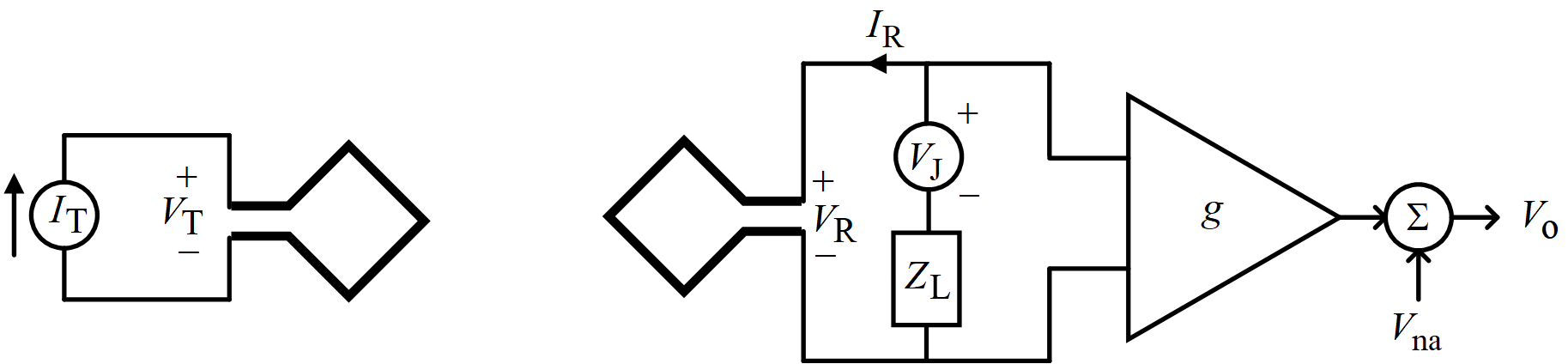}
\caption{Wireless link where the receiver front-end is colocated with the antenna: The transmit antenna is driven by a current source, and the receiver antenna is connected to a load impedance, $Z_{\mathrm{L}}$, whose voltage is measured by an infinite input impedance amplifier having voltage-gain, $g$; two sources of noise are amplifier noise, $V_{\mathrm{na}}$, and Johnson noise, $V_{\mathrm{J}}$, associated with the load resistance.}
\label{two_ant_system}
\end{figure}

Suppose, instead, that we eliminate the transmission line and put the amplifier within a fraction of a wavelength of the antenna as shown in Fig. \ref{two_ant_system}.

Any two antennas operating in a linear time-invariant medium are completely characterized by an impedance matrix which relates the voltages and currents on the two antenna ports, $\mathbf{V}(\omega) = \mathbf{Z}(\omega) \mathbf{I}(\omega)$ \cite{Ivrlac},\cite{Migliore},
\begin{equation} \label{Z_matrix_twobytwo}
\left[ \begin{array}{c} V_{\mathrm{T}}(\omega) \\ V_{\mathrm{R}}(\omega)
\end{array} \right]
= \left[  
\begin{array}{cc} 
Z_{\mathrm{T}}(\omega) & Z_{\mathrm{TR}}(\omega) \\
Z_{\mathrm{RT}}(\omega) & Z_{\mathrm{R}}(\omega)
\end{array} \right]
\left[ \begin{array}{c} I_{\mathrm{T}}(\omega) \\ I_{\mathrm{R}}(\omega)
\end{array} \right].
\end{equation}
Reciprocity implies that the impedance matrix is non-conjugate symmetric, $\mathbf{Z}^{\mathrm{T}} = \mathbf{Z}$, so $Z_{\mathrm{TR}} = Z_{\mathrm{RT}}$. Conservation of energy implies that the real part of the impedance matrix is non-negative definite.

The transmit antenna is driven by a current source, $I_{\mathrm{T}}$, and the receiver antenna is connected to a load impedance, $Z_{\mathrm{L}}$ (all quantities are functions of frequency). The voltage across the load impedance is measured by an infinite input-impedance amplifier having voltage gain, $g$. There are two sources of noise in the system: additive noise, $V_{\mathrm{na}}$, at the output of the amplifier having spectral density $N_{\mathrm{na}}$, and Johnson noise \cite{Bennett}, \cite{Franceschetti} associated with the load resistance, $V_{\mathrm{J}}$, having spectral density $2kT \mathrm{Re} Z_{\mathrm{L}}$, where $k$ is Boltzmann's constant, and $T$ is absolute temperature. \footnote{Our noise spectral densities are two-sided - hence the factor two rather than the more common factor of four in the Johnson formula.} Circuit theory yields the expression for the output voltage in terms of the transmitter current and the two noise sources,
\begin{equation} \label{Vout_Iin}
V_{\mathrm{o}} = g \left( \frac{Z_{\mathrm{L}}}{Z_{\mathrm{R}} + Z_{\mathrm{L}}}  \right) Z_{\mathrm{RT}} I_{\mathrm{T}} + V_{\mathrm{na}} + g \left( \frac{Z_{\mathrm{R}}}{Z_{\mathrm{R}} + Z_{\mathrm{L}}}  \right) V_{\mathrm{J}} .
\end{equation}
The resulting signal-to-noise ratio (as defined by a communication theorist) is
\begin{equation} \label{snr_2ant}
\mathrm{SNR}_{\mathrm{o}} = \frac{g^2 \left| \frac{Z_{\mathrm{L}}}{Z_{\mathrm{R}} + Z_{\mathrm{L}}}  \right|^2 | Z_{\mathrm{RT}} |^2 S_{\mathrm{I_{\mathrm{T}}}}  }{N_{\mathrm{na}} + g^2 \left| \frac{Z_{\mathrm{R}}}{Z_{\mathrm{R}} + Z_{\mathrm{L}}}  \right|^2 2kT \mathrm{Re} Z_{\mathrm{L}}} ,
\end{equation}
where $S_{\mathrm{I_{\mathrm{T}}}}$ is the spectral density of the transmitter current.

We now consider the SNR for two values of the load impedance.
\begin{itemize}
\item Open-circuit voltage measurement, $Z_{\mathrm{L}} = \infty$:
\begin{equation} \label{SNRoc}
    \mathrm{SNR}_{\mathrm{oc}} = \frac{g^2 |Z_{\mathrm{RT}} |^2 S_{\mathrm{I_{\mathrm{T}}}}}{N_{\mathrm{na}}} .
\end{equation}

\item Impedance matching, $Z_{\mathrm{L}} = Z_{\mathrm{R}}^*$:
\begin{equation} \label{SNRmatch}
    \mathrm{SNR}_{\mathrm{match}} = \frac{g^2 \left| \frac{Z_{\mathrm{R}}}{2 \mathrm{Re} Z_{\mathrm{R}}}  \right|^2 | Z_{\mathrm{RT}} |^2 S_{\mathrm{I_{\mathrm{T}}}}  }{N_{\mathrm{na}} + g^2 \left| \frac{Z_{ \mathrm{R}}}{2 \mathrm{Re} Z_{\mathrm{R}} }  \right|^2 2kT \mathrm{Re} Z_{\mathrm{R}}}  .
\end{equation}

\end{itemize}
The ratio of the respective SNRs is
\begin{equation} \label{snr_ratio}
\frac{\mathrm{SNR}_{\mathrm{oc}}}{\mathrm{SNR}_{\mathrm{match}}} = 4 \left|  \frac{\mathrm{Re} Z_{\mathrm{R}}}{Z_{\mathrm{R}}}  \right|^2 + \frac{g^2 2 k T \mathrm{Re} Z_{\mathrm{R}}}{N_{\mathrm{na}}} ,
\end{equation}
which may be either greater than or less than one. The second term in (\ref{snr_ratio}) represents the ratio of output Johnson noise voltage variance to the amplifier noise voltage variance; suppose, for example, that this ratio is less than one. If $\mathrm{Re} Z_{\mathrm{R}} \gg  \mathrm{Im} Z_{\mathrm{R}}$ then $\mathrm{SNR}_{\mathrm{oc}} > \mathrm{SNR}_{\mathrm{match}}$. The converse holds if  $\mathrm{Re} Z_{\mathrm{R}} \ll  \mathrm{Im} Z_{\mathrm{R}}$. In other words, impedance-matching to extract maximum power from the antenna does not always yield the optimum signal-to-noise ratio. Therefore maximum power extraction is not - and cannot - be of fundamental importance.

\subsection{Take-Away Points}
\begin{itemize}
    \item A transmission line requires termination in a resistor that is matched to the characteristic impedance of the line. To obtain maximum voltage on the terminating resistor for subsequent amplification, it is necessary to extract maximum power from the antenna. However this power only heats the resistor, and in no way contributes to the power of the signal at the output of the amplifier.

    \item The transmission line can be dispensed with provided the preamplifier is colocated with the antenna. One can connect the antenna port to a load impedance, and then measure the voltage on the resistor with a high impedance amplifier. Neither infinite load impedance, or matched impedance is universally superior. Conjugate matching, which extracts maximum power from the antenna, is not always optimal, and therefore is of no fundamental importance.
\end{itemize}

\section{What About Noise Figure?}

The concept of noise figure \cite{Friis} is pervasive in microwave engineering. For a two-port network, the noise factor is defined as the ratio of available signal to noise power (SNR) at the input of the network divided by the ratio of available signal to noise power at the output of the network. \footnote{When expressed in logarithmic form, noise factor becomes noise figure.} It is of critical importance to note that Friis' definition of SNR is distinct from the communication theorist's definition of SNR.

\subsection{Signal-to-Noise Ratio}
\begin{figure}
\centering
\includegraphics [scale=0.50] {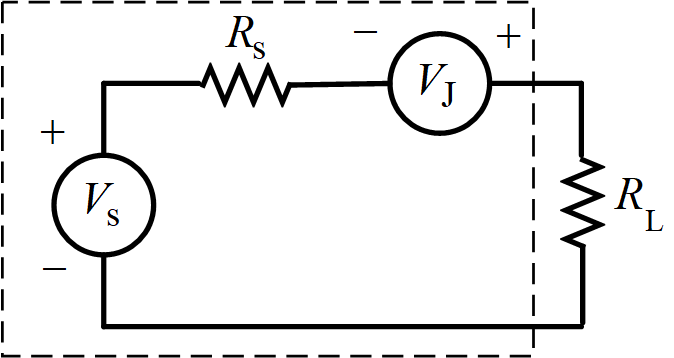}
\caption{Signal generator, comprising ideal voltage source, $V_{\mathrm{s}}$, in series with a resistor, $R_{\mathrm{s}}$ - a source of Johnson noise - connected to a load resistor, $R_{\mathrm{L}}$; the load resistor absorbs maximum power when matched to the internal resistance of the generator. The respective powers absorbed in the matched resistor due to the ideal voltage source and due to the Johnson noise of the source resistor constitute Friis' definition of available signal power and available noise power.}
\label{avail_power}
\end{figure}

Friis defines SNR as the ratio of ``available signal power" to ``available noise power". Consider a signal generator as in Fig. \ref{avail_power} comprising an ideal voltage source, $V_{\mathrm{s}}$, in series with a physical resistance, $R_{\mathrm{s}}$, where the only noise comprises the Johnson noise associated with the source resistance. The signal generator is connected to a load resistor, $R_{\mathrm{L}}$.
The available signal power is the maximum power associated with the generator's ideal voltage source which could be absorbed by the load resistor. Likewise the available noise power is the maximum Johnson noise power in the generator which could be absorbed by the load resistor. For a matching $R_{\mathrm{L}} = R_{\mathrm{s}}$, the signal and noise noise voltages are reduced by a factor of two due to the voltage divider effect, so the available signal power and the available noise power are:
\begin{equation}
    P_{\mathrm{s}} = \frac{|V_{\mathrm{s}}|^2}{4R_{\mathrm{s}}}, \ \ P_{\mathrm{n}} = \frac{\mathbb{E}|V_{\mathrm{J}}|^2}{4R_{\mathrm{s}}} = \frac{kT}{2} ,
\end{equation}
yielding a signal-to-noise ratio,
\begin{equation}
    \mathrm{SNR} = \frac{ |V_{\mathrm{s}}|^2 }{2 kTR_{ \mathrm{s}}}. 
\end{equation}

The output of a two-port network, for example an amplifier, comprises amplified desired signal and noise, where the noise arises both from the amplified input noise, and noise internal to the amplifier. The output of the network has some impedance, and therefore there is an available output signal power and an available output noise power, whose ratio is the output SNR.

\subsection{Gain}
The gain of the two-port network is defined as the ratio of the available output power to the available input power.

It is worth noting a comment by Friis: ``This is an unusual definition of gain since the gain of an amplifier is generally defined as the ratio of its output and input powers." ... ``Note that while the gain is independent of the impedance which the output circuit presents to the network, it does depend on the impedance of the signal generator."\cite{Friis}

In other words, ``gain" is not an intrinsic property of an amplifier.

\subsection{Noise Factor}
Noise factor is defined as the ratio of the input SNR to the output SNR,
\begin{equation} \label{noise_factor}
F = \frac{\mathrm{SNR}_{\mathrm{i}}}{\mathrm{SNR}_{\mathrm{o}}} .
\end{equation}

Again, a comment from Friis: ``The relationship between the noise figure and the degree of mismatch that exists between the network and its output and input circuits is important. Definition (4) shows clearly that the output circuit and hence its coupling with the network has no effect on the value of the noise figure. However, it also shows that the noise figure does depend on the degree of mismatch between the generator and the network since both $S$ and $N$ will vary with the magnitude of this mismatch."\cite{Friis}

Similar to ``gain", ``noise figure" is not an intrinsic property of an amplifier.

\subsection{Sample Calculation of Noise Factor}
\begin{figure}
\centering
\includegraphics [scale=.7] {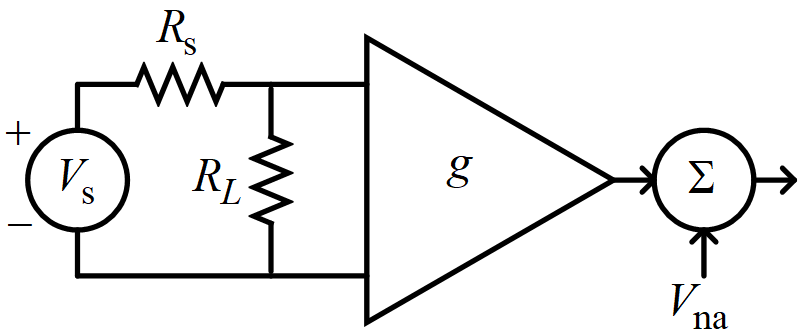}
\caption{Calculating the noise factor of a voltage amplifier: input voltage source $V_{\mathrm{s}}$ with internal resistance $R_{\mathrm{s}}$ is connected to a load-resistor, $R_{\mathrm{L}}$, whose voltage is measured by an infinite input impedance voltage amplifier having gain, $g$; noise sources comprise Johnson noise from the resistors, and internal amplifier noise, $V_{\mathrm{na}}$.}
\label{amp_noisefactor}
\end{figure}

Fig. \ref{amp_noisefactor} shows the signal generator of Fig. \ref{avail_power} connected to the voltage amplifier of Fig. \ref{two_ant_system}.
As before, the voltage gain is $g$, and the spectral density of the amplifier noise is $N_{\mathrm{na}}$. The amplifier has infinite input impedance, and output impedance equal to $R_{\mathrm{o}}$.

At the input, the available signal power, the available noise power, and the SNR are
\begin{equation} \label{sample_PN_in}
P_{\mathrm{si}} = \frac{|V_{\mathrm{s}}|^2}{4R_{\mathrm{s}}}, \  P_{\mathrm{ni}} = \frac{kT}{2}, \  
\mathrm{SNR_{\mathrm{i}}} = \frac{ |V_{\mathrm{s}}|^2 }{2 kTR_{ \mathrm{s}}}.
\end{equation}
The open-circuit signal-voltage at the output of the amplifier and the available output signal power are
\begin{equation} \label{out_voc}
V_{\mathrm{so}} = \frac{g V_{\mathrm{s}} R_{\mathrm{L}} }{ R_{\mathrm{s}} + R_{\mathrm{L}}}, \ \ P_{\mathrm{so}} =  \frac{g^2 |V_{\mathrm{s}}|^2}{4 R_{\mathrm{o}}} \left( \frac{ R_{\mathrm{L}} }{ R_{\mathrm{s}} + R_{\mathrm{L}}} \right)^2 .
\end{equation}
The Friis gain is
\begin{equation} \label{ex_gain}
G = \frac{P_{\mathrm{so}}}{P_{\mathrm{si}}} = \frac{ g^2 R_{\mathrm{s}} }{ R_{\mathrm{o}} } \left( \frac{ R_{\mathrm{L}} }{ R_{\mathrm{s}} + R_{\mathrm{L}}} \right)^2 .
\end{equation}

The Johnson noise at the input of the amplifier arises from the parallel combination of the source resistor and the load resistor. The available output noise power is
\begin{equation} \label{ex_out_noise}
P_{\mathrm{no}} = \left[ 2kT \left( \frac{  R_{\mathrm{s}} R_{\mathrm{L}} }{ R_{\mathrm{s}} +R_{\mathrm{L}} } \right) g^2 + N_{\mathrm{na}} \right] \, \frac{1}{4 R_{\mathrm{o}}},
\end{equation}
and the output SNR is
\begin{equation} \label{ex_SNR_o}
\mathrm{SNR}_{\mathrm{o}} = \frac{ g^2 |V_{\mathrm{s}}|^2 \cdot \left( \frac{ R_{\mathrm{L}} }{ R_{\mathrm{s}} + R_{\mathrm{L}}} \right)^2   }{ 2kT g^2 \left( \frac{  R_{\mathrm{s}} R_{\mathrm{L}} }{ R_{\mathrm{s}} +R_{\mathrm{L}} } \right) + N_{\mathrm{na}}  } .
\end{equation}

Finally the noise factor is
\begin{equation} \label{sample_F}
    F = \frac{\mathrm{SNR_{\mathrm{i}}} }{\mathrm{SNR}_{\mathrm{o}}  } = \left( \frac{ R_{\mathrm{s}} +R_{\mathrm{L}} }{ R_{\mathrm{L}} } \right) \left[ 1 + \frac{ N_{\mathrm{na}} }{ 2kT g^2 } \left( \frac{R_{\mathrm{s}} +R_{\mathrm{L}} }{R_{\mathrm{s}} R_{\mathrm{L}} } \right) \right] .
\end{equation}

Note that both the output SNR and the noise factor are independent of the output impedance of the amplifier, but they do depend on the source resistance and the input load resistance, in agreement with Friis' comment. The interplay of the source resistance and the load resistance is revealing:
\begin{itemize}
\item Gain: For fixed load resistance, gain is maximized by matching the source resistance to the input resistance,
\begin{equation} \label{G_max_Rs}
R_{\mathrm{s}} = R_{\mathrm{L}} \Rightarrow G = \frac{g^2 R_\mathrm{L}}{4 R_{\mathrm{o}}}.
\end{equation}
\item Gain: For fixed source resistance, gain is maximized for infinite load impedance,
\begin{equation} \label{G_max_Ri}
   R_{\mathrm{L}} = \infty \Rightarrow G = \frac{g^2 R_{\mathrm{s}}}{R_\mathrm{o}} . 
\end{equation}
\item Output SNR: For fixed load resistance, output SNR is maximized when the source resistance is equal to zero,
\begin{equation} \label{SNRo_max_Rs}
R_{\mathrm{s}}=0 \Rightarrow \mathrm{SNR}_{\mathrm{o}} = \frac{g^2 |V_{\mathrm{s}}|^2}{N_{\mathrm{na}}} ,
\end{equation}
\item Output SNR: For fixed source resistance, output SNR is maximized when the load resistance is infinite,
\begin{equation} \label{SNRo_ex_Ri}
R_{\mathrm{L}} = \infty \Rightarrow \mathrm{SNR}_{\mathrm{o}} = \frac{ g^2 |V_{\mathrm{s}}|^2   }{ 2kT R_{\mathrm{s}} g^2  + N_{\mathrm{na}}  } .
\end{equation}
\item Noise factor: For fixed load resistance, noise factor is minimized for 
\begin{equation} \label{F_ex_Rs}
R_{\mathrm{s}} = R_{\mathrm{L}} \sqrt{ \frac{N_{\mathrm{na}}}{2kTR_{\mathrm{L}} g^2 + N_{\mathrm{na}}} } \Rightarrow F = messy \, expression .
\end{equation}
\item Noise factor: For fixed source resistance, noise factor is minimized for
\begin{equation} \label{F_ex_RL}
R_{\mathrm{L}} = \infty \Rightarrow F = 1 + \frac{N_{\mathrm{na}}}{2kT R_{\mathrm{s}} g^2} .    
\end{equation}
\end{itemize}

With respect to all three criterion - gain, output SNR, and noise factor - for a given source resistance, performance is optimum when the load resistance is infinite. The observed behavior is in line with a comment by Friis: ``In amplifier input circuits a mismatch condition may be beneficial due to the fact that it may decrease the output noise more than the output signal."\cite{Friis}

Of the three amplifier properties discussed above, output SNR would be most relevant if the amplifier were directly connected to an A/D converter (here assuming that the input impedance of the A/D converter is much greater than the output impedance of the amplifier). The output SNR is equal to the ratio of the variance of the open-circuit output signal to the variance of the open-circuit output noise, and it coincides with the information-theoretic definition of SNR.

\subsection{The Real Utility of Noise Figure}
As defined by Friis, noise figure is a very general concept which can theoretically handle almost any connection of microwave devices. In practice, however, noise figure has a more restricted application as a standards specification for two-port devices having a common (50 Ohm) input and output impedance. The gain and the noise figure are defined for a standard temperature, and can easily be measured in the lab. The microwave engineer can connect devices together with any lengths of 50-Ohm coax cable and be assured that the overall system will function as predicted according to Friis' formula for the noise figure of a cascade of devices.

An impedance mismatch at the input of a device changes both the gain and the noise figure of the device from their nominal 50-Ohm values. In principle, the new gain and noise figure could be calculated, but in practice this cannot be done reliably because the calculation depends on the unknown details of the actual circuit. The only practical recourse is to take measurements in the lab.

It is worth noting that op amp noise specifications are invariably given in terms of effective input noise current and input noise voltage levels \cite{opamp_noisefigure}.

\subsection{Take-Away Points}
\begin{itemize}
\item{The concept of noise figure embodies no fundamental physical principle.}
\item{The definition ``available power at the input" implies neither that the amplifier must extract positive power at its input, nor that power must be extracted from the source.}
\item{The entire utility of noise figure is that it constitutes a standard definition of the gain and noisiness of an r.f. device when driven by a source of matched impedance, enabling 50-Ohm devices to be reliably connected together with 50-Ohm coax.}
\item{Noise figure doesn't tell us anything regarding how much power we must extract from a receiver antenna.}
\end{itemize}

\section{When is it Useful to Extract Maximum Power from a Receiver Antenna?} \label{Transformer}

We have established in previous sections that there is no fundamental mandate to extract maximum power from a receiver antenna, nor for that matter, to extract any power or current at all. Nevertheless, if the pre-amplifier has high input impedance with respect to the antenna's radiation impedance, then a matching network - in its simplest form a step-up transformer - can usefully provide noise-free voltage amplification of the antenna's open-circuit voltage prior to amplification. The use of a step-up transformer is well-known in the instrumentation literature \cite{Lekkala}, \cite{Achtenberg}. As we shall see, however, this activity has practical limitations.

\subsection{Step-Up Transformer}
\begin{figure}
\centering
\includegraphics [scale=.4] {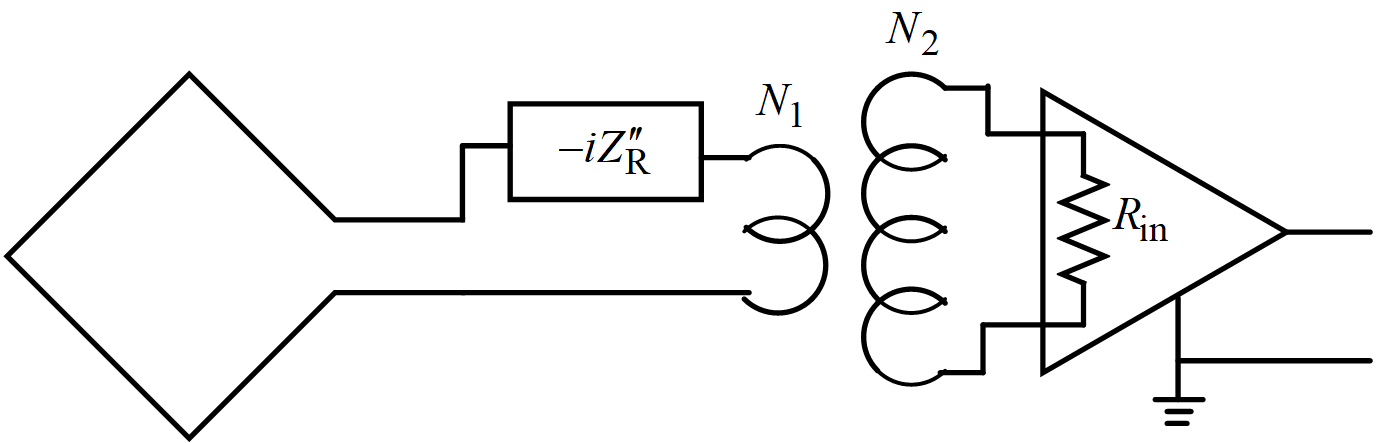}
\caption{When the input impedance of the amplifier is greater than the radiation impedance of the antenna, a step-up transformer can provide noise-free voltage gain equal to the turns-ratio, $N_2/N_1$, in advance of the amplifier; the primary winding sees a purely resistive impedance which can be matched to the input resistance, $R_{\mathrm{in}},$ of the voltage amplifier.}
\label{amp_transformer}
\end{figure}
In Fig. \ref{amp_transformer} the antenna is connected to the primary coil of an ideal transformer through a reactive device that cancels the imaginary part of the radiation impedance. The secondary coil is connected to a voltage amplifier having input resistance $R_{\mathrm{in}}$. The ideal transformer noiselessly amplifies the voltage on the primary coil by the factor equal to the turns-ratio, $N_2/N_1$. Too large a turns-ratio is counter-productive: the optimum is equal to the square-root of the impedance ratio,
\begin{equation} \label{opt_turns_ratio}
\frac{N_2}{N_1} = \sqrt{\frac{R_{\mathrm{in}}}{\mathrm{Re} Z_{\mathrm{R}}}} .
\end{equation}
With this turns-ratio, the primary coil presents an impedance of $\mathrm{Re} Z_{\mathrm{R}}$ to the antenna. Hence maximum power is extracted from the antenna.

\subsection{Practical Limitations} 
Suppose that an antenna has a self-impedance of 100 Ohms, and that the input impedance of the amplifier is $10^{12}$ Ohms. The impedance ratio is $10^{10}$, so theoretically we could use a transformer having a turns-ratio - and therefore voltage gain - of $10^5$. This is an impractical turns-ratio, but theoretically one could cascade several successive transformers. A further difficulty is that, in practice, r.f. transformers have both resistive losses and parasitic capacitance. The matching network that cancels the reactive self-impedance of the antenna is easy to implement at a single frequency, but increasingly difficult to realize as bandwidth increases. It would be challenging to achieve a voltage gain of $10^5$ over a wide band of frequencies.

\subsection{Take-Away Points}
\begin{itemize}
    \item A passive, lossless impedance matching network, such as a transformer, can theoretically provide noise-free voltage amplification in advance of an amplifier.
    \item In so doing, both current and power are extracted from the antenna, to a useful end.
    \item There are practical limitations on the actual voltage gain that can be realized over a wide band of frequencies.
\end{itemize}

\section{Does Impedance Matching for a Receiver Array Make Sense?}
Single-antenna communication links are becoming obsolete. Antenna arrays are the norm, and as the attention of 6G researchers increasingly focuses on the FR3 band ($\approx$ 7 - 24 GHz), these arrays will likely be tasked to function over extreme bandwidths. The notion of impedance matching for a receiver array is even more problematic than it is for a single receiver antenna.

\subsection{Circuit Theory}
\begin{figure}
\centering
\includegraphics [scale=.4] {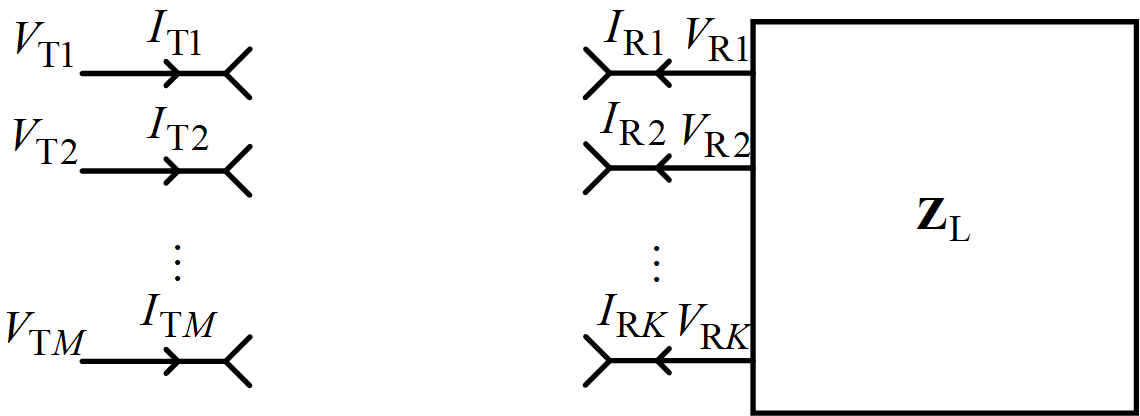}
\caption{MIMO link: receiver antennas are connected to a passive network.}
\label{MIMO_net}
\end{figure}
Fig. \ref{MIMO_net} illustrates a MIMO link comprising $M$ transmitter antennas and $K$ receiver antennas with the receiver antennas connected to a passive network having impedance $\mathbf{Z}_{\mathrm{L}}$. The transmitter and receiver arrays together are described by a partitioned impedance matrix,
\begin{equation} \label{MIMOZ} 
 \left[ \begin{array}{c} \mathbf{V}_{\mathrm{T}} \\ \mathbf{V}_{\mathrm{R}}
\end{array} \right]
= \left[  
\begin{array}{cc} 
\mathbf{Z}_{\mathrm{T}} & \mathbf{Z}_{\mathrm{TR}} \\
\mathbf{Z}_{\mathrm{RT}} & \mathbf{Z}_{\mathrm{R}}
\end{array} \right]
\left[ \begin{array}{c} \mathbf{I}_{\mathrm{T}} \\ \mathbf{I}_{\mathrm{R}}
\end{array} \right].
\end{equation}
Denote by $\mathbf{V}_{{\mathrm{R}}_{\mathrm{oc}}} = \mathbf{Z}_{\mathrm{RT}} \mathbf{I}_{\mathrm{T}}$ the open-circuit voltage that would be measured on the receiver antennas if no current were drawn, $\mathbf{I}_{\mathrm{R}} = 0$. The connection of the receiver array to the load network creates a matrix voltage-divider, such that
\begin{equation} \label{matrix_v_divider}
\mathbf{V}_{\mathrm{R}} = \mathbf{Z}_{\mathrm{L}} \left( \mathbf{Z}_{\mathrm{R}} + \mathbf{Z}_{\mathrm{L}} \right)^{-1} \mathbf{V}_{{\mathrm{R}}_{\mathrm{oc}}} .
\end{equation}

For any planar array - the type used for Massive MIMO for example - there is significant mutual coupling among the antennas. In general, the activity of extracting maximum power from one of the antennas occurs at the expense of the other antennas. This suggests the adoption of the criterion of maximum sum-power extraction, which entails conjugate-matching, $\mathbf{Z}_{\mathrm{L}} = \mathbf{Z}_{\mathrm{R}}^*$. Given our discussion of single-antenna links, however, there is no reason to expect this to be a universally optimal thing to do.

\subsection{Three Ways to Terminate a Receiver Array}
We now consider three choices for the load network and the resulting receiver antenna voltages:
\begin{itemize}
    \item Measure open-circuit voltage, $\mathbf{Z}_{\mathrm{L}} = \infty \, \mathbf{I}_K \ \Rightarrow \ \mathbf{V}_{\mathrm{R}} = \mathbf{V}_{{\mathrm{R}}_{\mathrm{oc}}}$, where $\mathbf{I}_K$ is the identity matrix;
    \item Conjugate-match each antenna, ignoring mutual coupling, 
    $\mathbf{Z}_{\mathrm{L}} = \mathbf{D}_{\mathbf{Z}_{\mathrm{R}}}^* \ \Rightarrow \ \mathbf{V}_{\mathrm{R}} =  \mathbf{D}_{\mathbf{Z}_{\mathrm{R}}}^* \left( \mathbf{Z}_{\mathrm{R}} + \mathbf{D}_{\mathbf{Z}_{\mathrm{R}}}^* \right)^{-1} \mathbf{V}_{{\mathrm{R}}_{\mathrm{oc}}} $, where $\mathbf{D}_{\mathbf{Z}_{\mathrm{R}}}$ is the diagonal matrix of antenna self-impedances;
    \item Conjugate-match, $ \mathbf{Z}_{\mathrm{L}} = \mathbf{Z}_{\mathrm{R}}^* \ \Rightarrow \ \mathrm{V}_{\mathrm{R}} = \frac{1}{2} \mathbf{Z}_{\mathrm{R}}^* \left( \mathrm{Re} \mathbf{Z}_{\mathrm{R}} \right)^{-1}  \mathbf{V}_{{\mathrm{R}}_{\mathrm{oc}}} $.

    \end{itemize}

    As previously noted, conjugate-matching of a single antenna over a wide frequency interval is itself a nontrivial activity, and hardly seems worth the trouble given that we are ignoring mutual coupling among the antennas. Furthermore, high bandwidth operation exacerbates mutual coupling: if the antennas are Nyquist-spaced at the high frequency end, then they are over-sampled at the low frequency end.

To perform full conjugate-matching for a receiver array, one has to create a passive network comprising capacitors, inductors, and transformers which duplicates all of the distributed electromagnetic phenomena of the antenna array itself over a wide band of frequencies. Practically speaking, this is a near-impossible task, which even if feasible would be uneconomical.

There is still another complication: complex-conjugation inherently entails time-reversal. The system $\mathbf{Z}_{\mathrm{R}}^* 
 (\omega)$ has an anti-causal impulse response. At a single frequency, or at a finite number of discrete frequencies, this is not a serious issue. But over a sufficiently wide band, causality becomes a difficult constraint.

Our analysis suggests that the activity of drawing currents from the antennas tends to exaggerate interaction among the antennas. This is certainly true for ideal ``invisible" antennas which disturb the ambient electromagnetic field only if current is drawn \cite{Ivrlac}. But actual patch antennas and array back-planes are associated with significant scattering of the ambient field even if the antenna ports are open-circuited.\footnote{For both invisible antennas and actual antennas, the impedance matrix still constitutes a complete and exact description of the electrical behavior of the array.} Nevertheless we conjecture that, as an alternative to current practice, it would be beneficial merely to measure the open-circuit voltages on the antenna ports.

\subsection{Take-Away Points}

\begin{itemize}

\item Impedance-matching is considerably more difficult for a receiver array than for a single antenna.

\item A potentially attractive alternative to drawing current from the antennas is simply to measure the open-circuit voltage on the antenna ports.
    
\end{itemize}

\section{Conclusions}

The central question posed in this paper is ``How much power must we extract from a receiver antenna to effect communications?". Our decisive answer is ``Subject to the laws of classical physics: None!".
\begin{itemize}
    \item The famous Shannon theory formula, $E_{\mathrm{b}}/N_0 > \ln 2$, is a purely mathematical result having no automatic physical significance.
    
    \item With existing technology it is entirely feasible to measure the open-circuit voltage on a receiver antenna port: no current drawn, no power drawn.
    
    \item Microwave engineering practice is largely dictated by the ubiquitous use of transmission lines which require terminating resistors which consume power that must come from the antenna; transmission lines can be dispensed with if the front-end electronics is colocated with the antenna.

    \item The Friis concept of noise figure constitutes a useful standards definition, but it does not tell us what we must do or not do.

    \item A low-impedance antenna can theoretically be matched to a high-impedance amplifier with a step-up transformer. Power is drawn from the antenna, but this power is usefully expended to provide noise-free voltage gain. Practically speaking, it can be difficult to realize this potential gain over a wide band of frequencies.

    \item The case for drawing power from an array of receiver antennas is even weaker than the case for drawing power from a single antenna. Measuring open-circuit voltages is not only feasible, but may have the added attraction of minimizing the effects of mutual coupling among the antennas. This conjecture requires experimental validation, however.
  
\end{itemize}

We have framed this paper entirely in the context of classical physics. In so doing we have ignored quantum mechanics which must be the ultimate arbiter of the central question posed in the paper. Our approach is justified by the fact that nobody designs a cellular wireless communication system according to the principles of quantum mechanics.

The fundamental question of whether or not power must be extracted from a receiver antenna, subject to the laws of classical physics, has evidently been of no interest either to wireless researchers or to microwave engineers. We have been unable to find any publications that attempt to answer this question either positively or negatively. We attribute this perplexing fact to an insufficient distinction being made between what is fundamentally possible or impossible to do, versus what is merely expedient to do.


%


\section{}
%



\ifCLASSOPTIONcaptionsoff
  \newpage
\fi


\begin{thebibliography}{1}

\bibitem{Shannon1}
C. E. Shannon, "A Mathematical Theory of Communication", \emph{The Bell System Technical Journal}, vol XXVII, no. 3, June 1948.

\bibitem{Shannon2}
Claude E. Shannon, "Communication in the Presence of Noise", \emph{Proc. I.R.E.}, vol 37, issue 1, Jan 1949.

\bibitem{Pierce}
J. R. Pierce, "Physical Sources of Noise", \emph{Proc. IRE}, pp. 601-608, May 1956. 

\bibitem{EDN_RF_opamp}
\emph{RF Design with Operational Amplifiers: Part I}, EDN, June 20, 2007.

\bibitem{TI_rf_opamp}
"OA-11 A Tutorial on Applying Op Amps to RF Applications", \emph{Texas Instruments Application Report}, revised April 2013.

\bibitem{AD8033}
\emph{AD8033/AD8034 Low Cost, 80 MHz FastFET Op Amps Data Sheet (Rev. D}, Analog Devices, Inc.

\bibitem{Inside_Out}
D. H. Sheingold (editor), \emph{Analog-Digital Conversion Handbook}, Nor-
wood, Massachusetts, Analog Devices, Inc., 1972.

\bibitem{Dolinar}
B. I. Erkmen, Baris, K. M. Birnbaum, B. E. Moision, and S. J. Dolinar, "The Dolinar receiver in an information theoretic framework", \emph{Proceedings of the SPIE}, Volume 8163, Aug. 2011.

\bibitem{Elliott}
R. S. Elliott, \emph{Antenna Theory and Design}, Wiley-IEEE Press, 2003.

\bibitem{Pozar}
D. M. Pozar, \emph{Microwave Engineering}, Wiley, 2nd edition, 1997.


\bibitem{Ivrlac}
M. T. Ivrlac and J. A. Nossek, “Toward a circuit theory of communi-
cation,” \emph{IEEE Transactions on Circuits and Systems I: Regular Papers},
vol. 57, no. 7, pp. 1663–1683, 2010

\bibitem{Migliore}
M. D. Migliore, "On electromagnetics and information theory", \emph{IEEE Trans. on Antennas and Propagation}, vol. 56, no. 10, pp. 3188-3200, 2008.

\bibitem{Bennett}
W. R. Bennett, \emph{Electrical Noise}, McGraw-Hill, 1960.

\bibitem{Franceschetti}
M. Franceschetti, \emph{Wave Theory of Information}, Cambridge University Press, 2017.

\bibitem{Friis}
H. T. Friis, "Noise Figures of Radio Receivers", \emph{Proc. of the IRE}, pp. 419 - 422, July 1944.

\bibitem{opamp_noisefigure}
"Op Amp Noise Figure: Don't be Mislead", \emph{Analog Devices, MT-052
Tutorial}, Oct 2008

\bibitem{Lekkala}
J. O. Lekkala and J. A. V. Malmivuo, "Noise reduction using a matching input transformer", \emph{J. Phys. E: Sci. Instrum.}, Vol. 14, l981.

\bibitem{Achtenberg}
K. Achtenberg, J. Mikołajczyk, C. Ciofi, G. Scandurra, and Z. Bielecki, "Transformer-based low frequency noise measurement system for the investigation of infrared detectors’ noise", \emph{Measurement}, Elsevier, 2021.




\end{thebibliography}
\end{document}